\documentclass[doublecol]{epl2}
\usepackage{amssymb}
\usepackage{amsmath}

\newcommand{\eqb}{\begin{eqnarray}}
\newcommand{\eqe}{\end{eqnarray}}

\newcommand{\bH}{{H}}
\newcommand{\bF}{{F}}

\newcommand{\bw}{\vect{w}}

\def\bes{\begin{subequations}}
\def\ees{\end{subequations}}

\newcommand{\PT}{{\cal PT}}

\newcommand{\diag}{\mbox{diag}}

\title{Discrete solitons in  ${\cal PT}$-symmetric lattices}
\shorttitle{Discrete solitons in  ${\cal PT}$-symmetric lattices}

\author{V. V. Konotop\inst{1} \and D. E. Pelinovsky\inst{2} \and D. A. Zezyulin\inst{1}}
\shortauthor{V. V. Konotop \etal}

\institute{
  \inst{1} Centro de F\'isica Te\'orica e Computacional and Departamento de F\'isica,
Faculdade de Ci\^encias, Universidade de Lisboa, Avenida Professor Gama Pinto 2, Lisboa 1649-003, Portugal\\
  \inst{2} Department of Mathematics and Statistics, McMaster University, Hamilton, Ontario, L8S 4K1, Canada
}
\pacs{63.20.Pw}{Localized modes}
\pacs{05.45.Yv}{Solitons}
\pacs{42.65.Wi}{Nonlinear waveguides}

\abstract{We prove existence of discrete solitons
in infinite  parity-time ($\PT$-) symmetric lattices by means
of analytical continuation from the anticontinuum limit.
The energy balance between dissipation
and gain  implies   that in the anticontinuum limit the solitons
are constructed from elementary $\PT$-symmetric blocks such as dimers,
quadrimers, or more general oligomers.
We consider in detail a chain of coupled dimers, analyze bifurcations of
discrete solitons from the anticontinuum limit and show that the solitons
are stable in a sufficiently large region of the lattice parameters.
The generalization  of the approach is
illustrated on two examples of networks of quadrimers, for which
stable discrete solitons are also found.}

\begin{document}

\maketitle

\section{Introduction}

{Energy localization in lattices is a fundamental topic. It
received its particular significance after the prediction of the
intrinsic localized modes~\cite{Takeno} and
the subsequent rigorous proof of the existence of such modes~\cite{McKay1994}
by analytical continuation from the anticontinuum limit when
the coupling of the nearest neighbors is weak. Nowadays the
topic is very well elaborated and numerous physical applications,
including nonlinear optics and ~\cite{disc_opt}
Bose-Einstein condensates~\cite{BEC1} have been thoroughly
studied.} One of the most popular model appearing in
description of these physical phenomena, which is also a
widely accepted testbed for mathematical analysis of the
anticoninuum limit, is the discrete nonlinear Schr\"odinger
equation (DNLS)~\cite{Surv_Tsir1}, also known as the discrete self-trapping
equation~\cite{Eilbeck}.

More recently,  particular attention was paid to the DNLS with
gain and losses. Such models naturally appear in the optical
context of arrays of amplifying and absorbing
waveguides~\cite{diss_waveguide}.
If gain and
losses  are adjusted to create the refractive index profile having
symmetric real and anti-symmetric imaginary parts~\cite{Muga}, such systems
have parity-time ($\PT$) symmetry and may have pure
real spectrum. Originally the idea about existence of pure real spectrum of
complex potentials obeying the $\PT$-symmetry  was introduced in~\cite{Bender}
questioning the fundamentals of the quantum mechanics. It turned
out however that the most direct applications of the $\PT$
symmetry today can be found in the discrete optics. Namely, in such
systems, and more specifically in two coupled waveguides (one with dissipation
and another with gain) the phenomenon was implemented experimentally~\cite{experiment2}.

Many detailed studies of $\PT$-symmetric DNLS equation were already
developed for lattices with a finite number of sites.
In particular, there were considered periodic oscillations in a system of two oscillators
(a {\em dimer})~\cite{Christ1}; stationary nonlinear modes for
four oscillators (a {\em quadrimer})~\cite{Kevr2011,ZK};
the relation between one- and two-dimensional finite $\PT$-symmetric
networks~\cite{ZK}, and detailed analysis of two-dimensional
plaquettes~\cite{plaquete}. The transition to the limit of an infinite number
of sites was investigated in~\cite{instability}.
It was shown that in this limit the $\PT$-symmetry breaking occurs at gain-loss
coefficient approaching zero. Stable discrete solitons  in the infinite
chain of $\PT$-symmetric DNLS oscillators with alternating coupling  were discovered
numerically  in~\cite{BinaryLattice}. However, the solitons obtained
in~\cite{BinaryLattice}, displayed oscillations and neither analytical
proof of the existence nor the number of possible families of solutions were clarified, so far.

In this Letter, we give an analytical proof of the existence of   discrete
solitons in $\PT$-symmetric DNLS lattices
with alternating coupling coefficients and classify different solution families
and their stability near the anticontinuum limit. In particular,
we report multistability of localized modes, that is, the
existence of two or more stable solutions with the same energy and
the same lattice parameters (but having different shapes).

{The DNLS equation with alternating coefficients of gain and loss can be
viewed is a discrete (tight-binding) limit of a continuous $\PT$-symmetric lattice.
Stable solitons in such systems have been found in the presence of only
linear~\cite{lin}, only nonlinear~\cite{nonlin}, and both linear and
nonlinear~\cite{lin-nonlin} $\PT$-lattices. The solutions
considered in this Letter can be viewed as discrete counterparts of the  mentioned solitons.}

\begin{figure}
\onefigure[width=0.75\columnwidth]{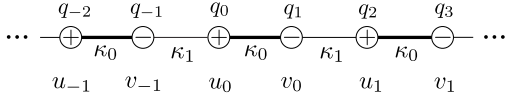} \caption{Schematic
presentation of the $\PT$-symmetric array of waveguides with gain,
``$+$'', and losses, ``$-$'', and  with the notations used in
the text.} \label{fig0}
\end{figure}

We consider the $\PT$-symmetric DNLS equation
\begin{eqnarray}
\label{dynam}
    i \frac{d q_n}{d t}
    + c_n(q_{n+1}-q_n)+c_{n+1}(q_{n-1}-q_n)  - g |q_n|^2q_n
    \nonumber \\
    +i(-1)^{n+1}\gamma q_n=0,
\end{eqnarray}
where the positive constants $c_n=\kappa_0$ for $n=2p$ and $c_n=\kappa_1$ for
$n=2p+1$ describe the two alternating couplings ($\kappa_0$ and
$\kappa_1$, with $\kappa_{0,1}>0$) between neighbor sites, and it is assumed that all odd
(even) sites have loss (gain) which is characterized by factor
$\gamma>0$ [see fig.~\ref{fig0}].
In the context of
optical applications our model describes an array of waveguides with
gain and losses. Then $q_n$ is
a dimensionless field in the waveguide $n$, and $t$ means the
propagation coordinate.

Let us first briefly address the most important features of the underlying
linear problem, which can be formally obtained from eq.~(\ref{dynam}) by setting $g=0$.
Identifying solutions of the linear problem in the form of Floquet-Bloch modes
$(q_{2n}, q_{2n+1}) = (u, v) e^{i k n - i(\kappa_0 + \kappa_1 + \mu)t}$, one can
recover that ${\PT}$ symmetry is unbroken on the infinite lattice if~\cite{BinaryLattice}
 \begin{equation}
 \label{eq:unbroken}
\bigl|\kappa_0 - \kappa_1\bigr| \geq \gamma.
  \end{equation}
In this case  for any real $k$ the corresponding eigenvalue $\mu$ is real.
More precisely $\mu^2$ lies in the interval
\begin{equation}
\label{eq:spectral-band}
(\kappa_0 - \kappa_1)^2 - \gamma^2 < \mu^2 < (\kappa_0+ \kappa_1)^2 - \gamma^2,
\end{equation}
i.e. $\mu$ belongs either to a positive or to a   negative spectrum band.
If inequality (\ref{eq:unbroken}) does not hold, we say that
$\PT$ symmetry is broken as there exist eigenvalues $\mu$ with nonzero imaginary parts.

Returning to the nonlinear problem, without loss of generality we impose $g =\pm 1$. Furthermore, by
analogy with the conservative DNLS equation (see e.g.~\cite{ABK}),
one can verify that there exists the symmetry reduction as
follows. If $q_n$ is a solution of (\ref{dynam}) for $g=1$, then
$(-1)^n \bar{q}_n e^{{-}2i(\kappa_0 + \kappa_1)t}$
is a solution of (\ref{dynam}) for $g=-1$. This
reduction allows us to restrict further consideration to the
case $g=1$ only.

\section{Anticontinuum limit}

We are concerned with the stationary
solutions, i.e. solutions whose dependence on time is given by
$q_n(t)\sim e^{-iEt}$, where $E$ is a constant which is termed
an energy (or a propagation constant in optical applications).
Then, in the case of the conservative DNLS equation, the anticontinuum
limit corresponds either to a coupling between the two neighbor sites
tending to zero or to the energy tending to the infinity, the two
limits being equivalent, i.e. mapped to each other by simple transformation (see
e.g.~\cite{ABK}). The same is true for the model (\ref{dynam}), however
with two important changes.

First, {in the presence of the dissipative
term $i(-1)^n\gamma q_n$} the elementary cell of
the DNLS equation (\ref{dynam}) is composed of
two sites, one with gain and the other one with loss
(even if $\kappa_0=\kappa_1$). {Therefore,} the
anticontinuum limit must be formulated in terms of dimers, rather than single sites. Thus, the anticontinuum limits can be identified as either $\kappa_0$ or
$\kappa_1$ to be small enough.
Without loss of
generality, we fix $\kappa_0= 1$. Then the anticontinuum limit corresponds to
$\kappa_1 = 0$.

In physical applications, both coupling constants $\kappa_0$ and
$\kappa_1$ are usually fixed. Then the anticontinuum limit can be realized
at the limit of large energy $E$. However, and this
is the second distinction from the conservative DNLS equation,
although the small parameter can be obtained in the
limit of large energy, $E$ cannot be scaled out from the
main equations if $\gamma \neq 0$.

Using the above considerations, it is convenient to rewrite the main model
(\ref{dynam}) in terms of variables
\begin{eqnarray}
\left(\!\begin{array}{c} q_{2n}(t) \\ q_{2n+1}(t) \end{array}\! \right)
=\left(\!\begin{array}{c} u_n\\
v_n\end{array} \!\right)    e^{-i(\kappa_0 + \kappa_1 + \mu) t},
 \end{eqnarray}
where $\mu$ is a constant and we assume that $u_n$ and $v_n$ do
not depend on $t$ and satisfy zero boundary conditions: $u_n,v_n\to 0$ as $n\to\pm\infty$. Then for $\kappa_0 =
1$, $\kappa_1 = \epsilon$, and $g =1$, the main model can be
rewritten in the matrix form for $\bw_n = (u_n,v_n)^T$ (``$T$'' stands
for the matrix transposition):
\begin{eqnarray}
\label{main}
{\bH} \bw_n +
\epsilon\left(\sigma_-\bw_{n-1}+\sigma_+\bw_{n+1}\right) = {\bF}(\bw_n) \bw_n,
\end{eqnarray}
where
\begin{eqnarray*}
{\bH} = \left(\begin{array}{cc} \mu-i\gamma& 1\\
1& \mu+i\gamma \end{array} \right), \\
\sigma_- = \left(\begin{array}{cc} 0&1\\ 0 & 0 \end{array}
\right), \quad \sigma_+ = \left(\begin{array}{cc} 0&0 \\1 &0
\end{array} \right),
\end{eqnarray*}
and ${\bF}(\bw_n) = {\rm diag}(|u_n|^2,|v_n|^2)$.

\section{Single-dimer state}
First, we address the simplest case when in the anticontinuum limit
$\epsilon=0$, only one central dimer is excited, i.e. $\bw_0\neq 0$,
whereas $\bw_n = {0}$ for $n \neq 0$. Then for any $n\ne 0$
eq.~(\ref{main}) is automatically  satisfied.  For the central dimer, i.e.
at $n=0$, we assume that the following $\PT$ symmetry reduction holds:
$u_0=\bar{v}_0$ and arrive at the following equation~\cite{Christ1}:
\begin{equation}
\label{lim-prob} (\mu - i \gamma) u_0 + \bar{u}_0 = |u_0|^2 u_0.
\end{equation}
The latter equation has two branches of solutions, which exist for all $\gamma \in [0,1)$:
\begin{eqnarray}
\label{dimer} u_0^{\pm} = A_{\pm} e^{i\varphi_{\pm}},
\end{eqnarray}
where $A_{\pm}^2 = \mu \pm \sqrt{1 - \gamma^2}$, $\sin(2
\varphi_{\pm}) = -\gamma$, and $\cos(2\varphi_{\pm}) = \pm \sqrt{1
- \gamma^2}$. The branches $u_0^{\mp}$ exists for $\mu > \mu_\pm $
where $\mu_{\pm}= \pm \sqrt{1 - \gamma^2}$.
The physical difference between these branches becomes
evident if we introduce the linear momentum
$p_n=2$Re$(\bar{q}_n q_{n+1})$ and the current
$j_n=2$Im$(\bar{q}_n q_{n+1})$ per unit cell; respectively $p=\sum_n
p_n$ are $j=\sum_n j_n$ are the total momentum and current carried
out by the solution.  Then,  branch   $u_0^+$ (branch $u_0^-$)  corresponds
to the linear momentum and current, between the two sites of the dimer,
having the same (opposite) directions.
{Note that eq.~(\ref{lim-prob}) coincides
with the equation for  the stationary solutions of the parametrically driven NLS equation~\cite{Barash_EPL,Barash_PRE}.}


Looking for continuation of the solution (\ref{dimer}) from the
anticontinuum limit (i.e. from $\epsilon=0$ to $\epsilon>0$) it
is natural to suppose that for $\epsilon>0$  the solitons   also
obey the symmetry reduction  on the whole infinite lattice, i.e.
\begin{equation}
\label{reduction1} u_n = \bar{v}_{-n}, \quad v_n = \bar{u}_{-n},
 \quad n =0, \pm 1, \pm2, \ldots.
\end{equation}
It allows one to restrict the consideration only to $n\geq 0$. At the
central dimer, $\bw_0$, one can introduce the real coordinates
$(a_0,b_0)$ such that $u_0 = \bar{v}_0 = a_0 + i b_0$ and rewrite (\ref{main}) for $n=0$ as follows
\begin{equation}
\label{problem-1-real}
\left\{ \begin{array}{l} \mu a_0 + \gamma b_0 + a_0 + \epsilon {\rm Re}(u_1) = (a_0^2 + b_0^2) a_0, \\
\mu b_0 - \gamma a_0 - b_0 - \epsilon {\rm Im}(u_1) = (a_0^2 +
b_0^2) b_0. \end{array} \right.
\end{equation}
For  $n\geq 1$, we still
use the complex-valued coordinates:
\begin{equation}
\label{problem-2}
\left\{ \begin{array}{l}  (\mu - i \gamma) u_n + v_n + \epsilon v_{n-1} = |u_n|^2 u_n, \\
(\mu + i \gamma) v_n + u_n + \epsilon u_{n+1} = |v_n|^2 v_n.
\end{array} \right. 
\end{equation}
Now the system (\ref{problem-1-real})--(\ref{problem-2}) is smooth
with respect to parameter $\epsilon$ and the solution vector. At
$\epsilon = 0$, we have the limiting solution: $u_n = v_n = 0$ for
all $n=1,2, \ldots$, while $a_0$ and $b_0$ are given by one of
the two possible solutions (\ref{dimer}) for $\gamma \in [0,1)$
and $\mu > \mu_{\pm}$. To apply the implicit
function theorem arguments, we need to show that the
Jacobian operator of the system
(\ref{problem-1-real})--(\ref{problem-2}) at $\epsilon = 0$ is
invertible at the limiting solution. Furthermore, the solution can
be analytically continued from the anticontinuum limit until a
critical value $\epsilon_{cr} > 0$ for which the Jacobian operator
becomes non-invertible.

In the case of the conservative DNLS equation ($\gamma=0$)
rigorous estimates for $\epsilon_{cr}$ can be obtained
analytically~\cite{McKay1994,ABK}. Due to mathematical constrains
such estimates are typically lower than the practically achievable
values of  $\epsilon_{cr}$ for which the localized solutions exist, on
the one hand, and on the other hand require more elaborated
analytical study. Therefore, here we restrict the consideration
only to the proof that the analytical continuation is possible and
study the continuation numerically.

For $\epsilon=0$, the lattice consists of a set of uncoupled
dimers. For  $n=1,2,\ldots$, the limiting Jacobian operator of the system
(\ref{problem-2}) is nothing but $\bH$ and thus is invertible for
$\mu \neq \mu_{\pm}$ since $\det(\bH) = \mu^2 +
\gamma^2 - 1$.

At the central dimer $n = 0$, the limiting Jacobian operator of the
system (\ref{problem-1-real}) is given by the $2 \times 2$ matrix:
\begin{eqnarray}
J_0 = \left( \begin{array}{cc} -2 a_0^2 - \gamma b_0/a_0 & \gamma - 2 a_0 b_0 \\
-\gamma - 2 a_0 b_0 & -2 b_0^2 + \gamma a_0/b_0 \end{array}
\right),
\end{eqnarray}
where   eq.~(\ref{lim-prob}) has been used. The
matrix $J_0$ is invertible if $a_0 b_0 \neq 0$ and $a_0^2 \neq
b_0^2$. This gives the constraints $A_{\pm} \neq 0$ and $\cos(2
\varphi_{\pm}) \neq 0$ in the limiting solution (\ref{dimer}), or
equivalently, $A^2_{\pm} \neq \{ 0, \mu\}$. The constraints
are satisfied for any $\gamma \in [0,1)$ and $\mu \neq \mu_{\pm}$.
Hence, for any $\gamma$ and $\mu$ that satisfy the found invertibility conditions,
solutions  $u_0^{\pm}$ of the dimer problem {give birth} to two branches of
localized discrete solitons on the infinite $\PT$-symmetric lattice.  These branches,
which will be respectively denoted as $\Gamma^{(\pm)}$ [see fig.~\ref{fig05}, below],
are parameterized by $\epsilon$, they  persist at least for all small
$\epsilon$, and   for  small $\epsilon$  the solitons from  the branches
$\Gamma^{(\pm)}$ are  nearly localized at the central dimer $\bw_0$.

\section{Multi-dimer states}

One can also consider the case  when the solution in the anticontinuum
limit consists of several excited dimers.
Say, for the case of two dimers, one can consider branches
$\Gamma^{(+, +)}$ or $\Gamma^{(-, -)}$, which at $\epsilon=0$
correspond to two dimers occupying two
{consecutive}  central cells $n=0$ and $n=1$. More complex
configurations consisting of $N$ excited dimers can also be
continued from the anticontinuum limit. Even more generally,
there exist branches like $\Gamma^{(+, 0, +)}$,  $\Gamma^{(-, 0, -)}$,
which  in the anticontinuum limit correspond to two  dimers placed at
$n=-1$ and $n=1$ separated with an ``empty'' dimer at $n=0$ (i.e. $\bw_0=0$
for $\epsilon =0$). However,  the continuation is not possible for arbitrary
choice of $N$ central dimers. If we consider the  sequence
$\vect{\alpha} = (\alpha_1, \alpha_2, \ldots, \alpha_{N})$ consisting
of $N$ symbols $\alpha_{1, \ldots, N}\in\{+, -, 0\}$, then existence
of the branch $\Gamma^{\vect{\alpha}}$ is only possible provided
that  $\alpha_p=\alpha_{N+1-p}$ for $p=1,...,N$. The latter requirement
ensures that the   configuration, which is chosen to be continued from
$\epsilon=0$ to $\epsilon>0$,   obeys a $\PT$ symmetry reduction (\ref{reduction1}).
For example,  for $N=3$, the family $\Gamma^{(-, +, +)}$ can not be continued
from the anticontinuum limit.  However the family $\Gamma^{(-, +, -)}$ does
persist for small $\epsilon$. Invertibility of the  Jacobian
operators corresponding to such multi-dimer solutions can be proven
using similar technique  as for the case $N=1$ considered above.
{We note that the relevance of the symmetry properties for
possibilities of continuation of the branches of soliton solutions
is similar to the case of parametricalally driven NLS systems~\cite{Barash_PRE}.}

\section{Stability in the anticontinuum limit}
We shall also address linear  stability of solitons belonging to
the branches  $\Gamma^{(\pm)}$  bifurcating from the one-dimer states
in the anticontinuum limit. For $\epsilon=0$ and $n\ne 0$,
the dimers are decoupled and the stability of the zero solution
is determined by the spectrum of the matrix $H$ which has real eigenvalues
if $\gamma<1$. Hence the zero solution for $n\neq 0$ is stable.
Passing from $\epsilon=0$ to $\epsilon>0$, eigenvalues  $\lambda$
group together into bands of continuous spectrum.  For small positive $\epsilon$,
these bands are situated in the neutrally stable imaginary axis, and
they are separated from each other and from zero if $\gamma \in [0,1)$,
$\mu \neq \mu_{\pm}$, and $\mu \neq 0$.

Thus, to ensure stability of the single-dimer soliton,
we have to address only the stability of the central dimer $\bw_0$.
Considering a perturbed solution $\bw_0+
\vect{\psi}_0 e^{\lambda t} + \overline{\vect{\psi}}_1e^{\bar\lambda t}$
and linearizing the equation  with respect to $\vect{\psi}_{0,1}$,
for $\epsilon = 0$ we obtain the eigenvalue problem
\begin{eqnarray}
\label{eigen}
\left(\!%
\begin{array}{cc}%
L_0 & L_1\\
-\bar{L}_1 & -\bar{L}_0
\end{array}\!\right)\left(\!
\begin{array}{c}{\vect{\psi}}_0\\%
{\vect{\psi}}_1\end{array}\!\right) =
i\lambda%
\left(\!
\begin{array}{c}{\vect{\psi}}_0\\%
{\vect{\psi}}_1\end{array}
\!\right),
\end{eqnarray}
where $L_0 = 2\,\textrm{diag}(|u_0^\pm|^2, |u_0^\pm|^2) - H$ and
$L_1 = \textrm{diag}((u_0^\pm)^2, (\bar{u}^\pm_0)^2)$, where
$u_0^{\pm}$ are defined by eq.~(\ref{dimer}) for branches $\Gamma^{(\pm)}$.
Because the nonlinear system (\ref{main}) admits gauge invariance,
$\lambda = 0$ is a double eigenvalue of the
eigenvalue problem (\ref{eigen}). As a result,
the characteristic polynomial $D(\lambda)$
can be factorized by $\lambda^2$ and reads as (see also \cite{Kevr2011}):
$
D(\lambda) = \lambda^2 \left( \lambda^2 +
8 \sqrt{1-\gamma^2} (\sqrt{1-\gamma^2} \pm \mu/2) \right),
$
where eq.~(\ref{dimer}) has been used.
This expression shows that  for $\epsilon=0$ and $\epsilon \ll
1$ the solitons from branch $\Gamma^{(+)}$ are stable for any
$\mu > \mu_-$ and $\gamma \in [0,1)$.  Solitons of the branch  $\Gamma^{(-)}$
are stable  for $\epsilon=0$ and $\epsilon \ll 1$ only if
$\mu_+ < \mu < 2 \mu_+$  and unstable with a positive
eigenvalue for $\mu > 2 \mu_+$.

\section{Numerical results}

\begin{figure}[h]
\onefigure[width=\columnwidth]{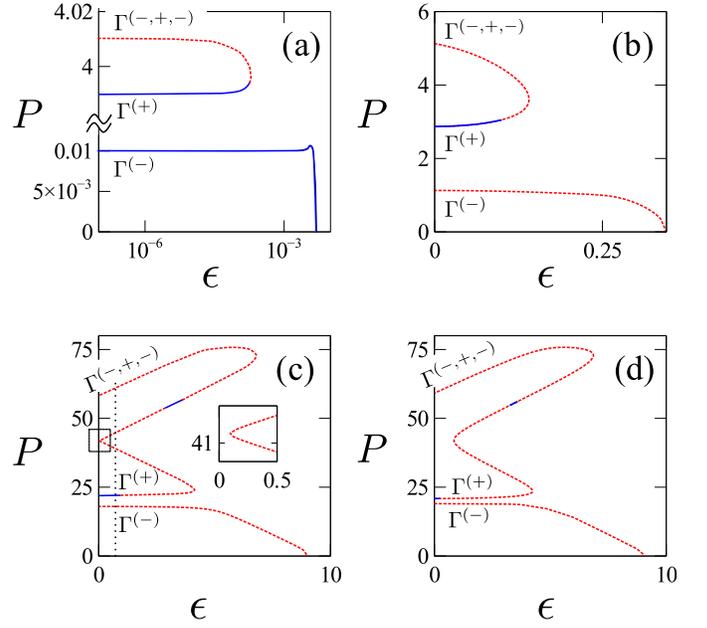} \caption{$P$
{\em vs.} $\epsilon$. Panels in the top (bottom) rows
correspond to $\mu=1$ ($\mu=10$). Panels in the
left (right) column correspond to $\gamma=0.1$ ($\gamma=0.9$).
Stable (unstable) solitons are shown by solid blue (dotted red) lines. Notice that panel
(a) has logarithmic scale in the horizontal axis and  broken
vertical axis. {Vertical dotted line in panel~(c) corresponds to $\epsilon=0.8$; see also fig.~\ref{fig06}.}} \label{fig05}
\end{figure}

Turning now to the numerical study of the discrete solitons in the infinite lattice,
we have computed bifurcations of families $\Gamma^{(\pm)}$ from
the anticontinuum limit $\epsilon=0$, considered their continuations
to the domain $\epsilon>0$, and examined stability of the found solitons.
The results are conveniently visualized in the plane ($P,\epsilon$),
where $P = \sum_{n} (|u_n|^2 + |v_n|^2)$, which in optics corresponds to the total energy flow.
In fig.~\ref{fig05} we show the results for different $\mu$ and $\gamma$.
We recall that the branch $\Gamma^{(+)}$ ($\Gamma^{(-)}$) is found by means of
continuation starting from the dimer solution $u_0^+$ ($u_0^-$)  given by  eq.~(\ref{dimer}).
We tested several values of $\mu$ and $\gamma$  and in all cases numerical
results for  stability of the families $\Gamma^{\pm}$    for small $\epsilon$
were in agreement with the above linear stability analysis. For example,
branch $\Gamma^{(-)}$ is stable for
$\mu=1$ and $\gamma=0.1$, but is  unstable for any other considered
values of $\mu$ and $\gamma$ on fig.~\ref{fig05}.

At a certain value of $\epsilon = \epsilon_{0}$,
the norm of the solitons belonging to the branch $\Gamma^{(-)}$ vanishes, i.e. $P\to 0$.
Since this is the linear limit,  at the point $P=0$  the parameters
obey the   relation  $\mu^2 = (1+\epsilon_{0})^2-\gamma^2$, which means that the solution branch
ends up at (or bifurcate from) the edge of the linear spectrum
[see eq.~(\ref{eq:spectral-band})]. Respectively,
$\epsilon_{0} = \sqrt{\mu^2+\gamma^2} - 1$.

\begin{figure}[h]
\onefigure[width=\columnwidth]{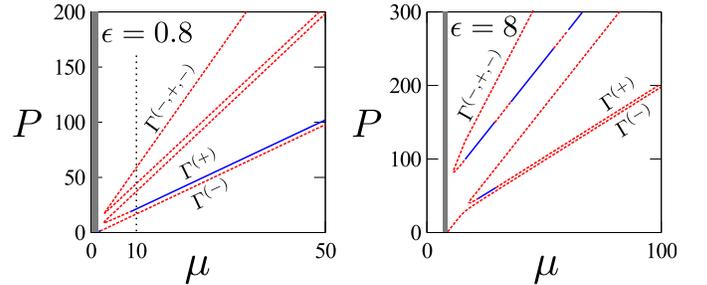} \caption{$P$
{\em vs} $\mu$ for $\gamma=0.1$ and different $\epsilon$.
Vertical shadowed domains show the bands of the linear spectrum.
{Vertical dotted line in the left  panel  corresponds to $\mu =10$; see also fig.~\ref{fig05}.}} \label{fig06}
\end{figure}

Bifurcation  of the discrete solitons from  the edge of the linear spectrum
becomes particularly evident if we employ representation  on the plane $(P,\mu)$,
which is to be  obtained  for fixed $\gamma$ and $\epsilon$. Then,
as shown in fig.~\ref{fig06},  the found discrete solitons   constitute continuous families,
which is a frequent feature of nonlinear  $\PT$-symmetric systems~\cite{ZK}.
{We notice that in the context of
 a parametrically driven NLS system, the connection
 of the soliton branch with the continuous spectrum was reported
 in~\cite{Barash_EPL}.}

\begin{figure}[h]
\onefigure[width=0.9\columnwidth]{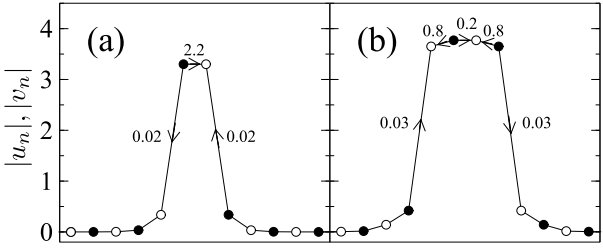}%
\caption{Amplitude   and currents for the unstable  soliton  at
$\epsilon=1$, $P \approx 22$ [panel (a)] and  for stable soliton  at $\epsilon =
3.32$,   $P\approx 56$ [panel (b)]. For both shown solitons $\mu=10$ and
$\gamma=0.1$. Filled and empty circles correspond to sites with gain (i.e. $u_n$)
and losses (i.e. $v_n$), respectively. Arrows show directions and amplitudes of the largest
currents in the system.}
\label{fig07}
\end{figure}

The branch $\Gamma^{(+)}$ is  stable for sufficiently small $\epsilon$  for all
considered $\mu$ and $\gamma$ (in agreement with the linear stability analysis).
For $\mu=10$ [see fig.~\ref{fig05}(c)-(d)] solitons of the family $\Gamma^{(+)}$
lose stability (the part of the continuous spectrum leaves the imaginary axis and
becomes unstable) at $\epsilon = 1-\gamma$,  i.e. at the $\PT$ symmetry
breaking bifurcation [see eq.~(\ref{eq:unbroken})].
{At $\epsilon = 1+\gamma$,  the $\PT$ symmetry is restored [see eq.~(\ref{eq:unbroken})],
but} the branch $\Gamma^{(+)}$ does not become stable, because isolated unstable eigenvalues
persist in the spectrum of linearization. Altogether, the branch  $\Gamma^{(+)}$
displays ``snaking'' behavior and has
several alternating domains of stability and instability. Finally,
branch $\Gamma^{(+)}$ returns to the anticontinuum limit by means
of coalescing  with the branch $\Gamma^{(-, +, -)}$ bifurcating from
the three-dimer state in the anticontinuum limit.

For
large $\mu$, e.g. $\mu=10$ in fig.~\ref{fig05}(c)--(d),
branches  $\Gamma^{(\pm)}$  can be continued  into the region   $\epsilon \in (1-\gamma, 1+\gamma)$,
where   $\PT$ symmetry is broken. In particular, solitons exist at $\epsilon = 1$, i.e.
$\kappa_0=\kappa_1$,  [fig.~\ref{fig07}(a)];  such solitons, however, are
unstable. A stable soliton  is  shown in fig.~\ref{fig07}(b).

\section{Generalizations}
\begin{figure}
\onefigure[width=0.95\columnwidth]{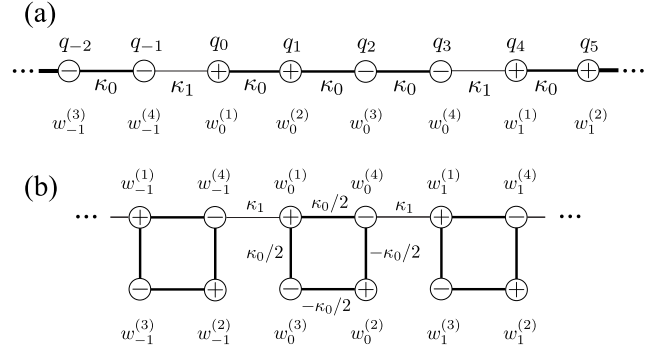} \caption{Two
examples of $\PT$-symmetric  networks, which consists of a set of
disconnected quadrimers in the anticontinuum limit.} \label{fig08}
\end{figure}

The developed approach can be applied to the case when the elementary cell of a network is
a more complex $\PT$-symmetric cluster (than the dimer). To illustrate this, we
now briefly address the anticontinuum limit for two networks of quadrimers, i.e. clusters of four sites $\bw_n=(w_n^{(1)},w_n^{(2)},w_n^{(3)},w_n^{(4)})^T$, whose examples are shown in
fig.~\ref{fig08}. To describe the network on fig.~\ref{fig08}(a),
we can still adopt  eq.~(\ref{main}), where
\begin{eqnarray}
\label{net1}
{H} = \left(\!
\begin{array}{cccc}
\mu-i\gamma & 1 & 0 & 0
\\
1 & \mu-i\gamma & 1 &0
\\
0 & 1 & \mu+i\gamma & 1
\\
0 & 0 & 1 & \mu+i\gamma
\end{array}
\!\right),
\end{eqnarray}
the nonlinearity is given by
$$
{F}(\bw_n) = \diag\left(|w_n^{(1)}|^2,|w_n^{(2)}|^2,|w_n^{(3)}|^2,|w_n^{(4)}|^2\right),
$$
and $\sigma_\pm$  are now $4\times 4$ matrices whose only {nonzero} elements
are $(\sigma_-)_{14}=(\sigma_+)_{41}=\epsilon$. The matrix $H$ is invertible
unless $\mu^2= \frac 32-\gamma^2 \pm \frac 12 \sqrt{5-16\gamma^2}$.

In the anticontinuum limit, defined by $\epsilon=0$, the
network shown in fig.~\ref{fig08} (a) consist  of a set of
disconnected $\PT$-symmetric quadrimers.
Here we consider the simplest case, when at $\epsilon=0$
only one central quadrimer is excited, i.e. $\bw_n=0$ for $n\ne 0$,
and look for continuation of this solution to $\epsilon>0$.

To prove the possibility of analytical continuation as above,
we concentrate on $\PT$-symmetric solutions (i.e. obeying the symmetry
$w_n^{(1)} = {\bar w_{-n}^{(4)}}$ and  $w_n^{(2)} = { \bar w_{-n}^{(3)}}$).
This allows us to restrict the consideration to the semi-infinite matrices
with $n\geq 0$. Moreover the invertibility of $H$
for $\mu^2 \neq \frac 32-\gamma^2 \pm \frac 12 \sqrt{5-16\gamma^2}$ ensures
the continuation provided the Jacobian matrix for the central quadrimer is invertible.

For $n=0$ and $\epsilon=0$,  the central quadrimer obeys  the
system of four equations [see eq.~(\ref{main})]
\begin{equation}
    \label{eq:sysq0}
    H\bw_0=F(\bw_0)\bw_0,
\end{equation}
under the symmetry: $w_0^{(1)} = {\bar w_0^{(4)}}$ and  $w_0^{(2)} = { \bar w_0^{(3)}}$.
While the nonlinear system (\ref{eq:sysq0}) generally does not admit an explicit analytical solution
[in contrast to the dimer case (\ref{dimer})], its properties are well studied.
In particular, families of its nonlinear modes, bifurcation diagrams, and some exact
solutions have been reported~\cite{Kevr2011, ZK,plaquete}.

\begin{figure}[h]
\onefigure[width=0.95\columnwidth]{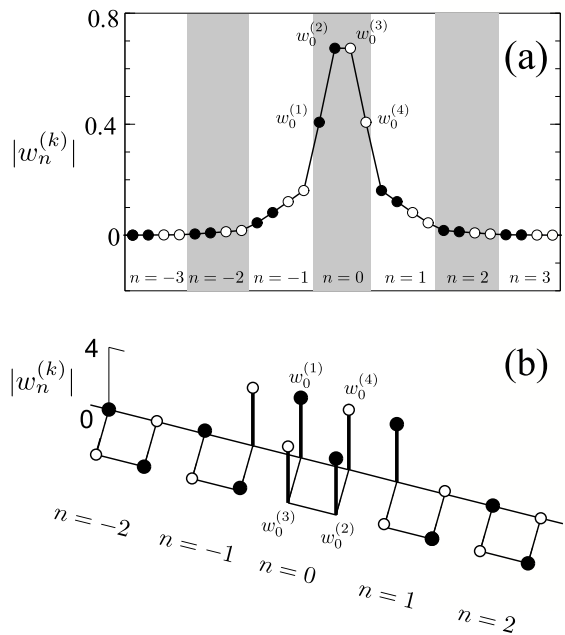}
\caption{(a)   Stable soliton for the network in
fig.~\ref{fig08}(a) at $\epsilon=0.5$, $\gamma=0.25$ and  $\mu=2$.
(b) Stable soliton for the network in
fig.~\ref{fig08}(b) at $\epsilon=1.6$,
$\gamma=0.1$ and $\mu=10$.
Filled (empty) circles correspond to sites with gain (losses).}
\label{fig10}
\end{figure}

The network in fig.~\ref{fig08}(a) is characterized by two types of the $\PT$ symmetry,
the {\em local} and {\em global} ones. {We say that the lattice is locally
$\PT$-symmetric if the system (\ref{eq:sysq0}) is $\PT$-symmetric in the limit $\epsilon = 0$.
On the other hand, we say that the lattice is globally $\PT$-symmetric if }
the infinite  network (\ref{main})
with the matrix $H$ in (\ref{net1}) is $\PT$-symmetric for $\epsilon \neq 0$.
The network in fig.~\ref{fig08}(a) consists of quadrimers which have unbroken
local $\PT$ symmetry, at least for small $\gamma$. For $\epsilon>0$ the infinite
system has unbroken global  $\PT$ symmetry allowing for stable discrete solitons.
An example of a stable discrete soliton for this network is shown on fig.~\ref{fig10}(a).

We shall now consider the network,  which consists of clusters whose local $\PT$ symmetry
is broken.  However, proper choice of the coupling $\epsilon>0$ makes the infinite network
to possess unbroken global $\PT$ symmetry. An example of such network is
presented in fig.~\ref{fig08}(b). For this network, we can still work with  eq.~(\ref{main}), where
\begin{eqnarray}
{\bH}= \left(\!
\begin{array}{cccc}
\mu-i\gamma & 0 & 1/2 & 1/2
\\
0 & \mu-i\gamma & -1/2 & -1/2
\\
1/2 & -1/2 & \mu+i\gamma & 0
\\
1/2 & -1/2 & 0 & \mu+i\gamma
\end{array}
\!\right). \label{net2}
\end{eqnarray}
Local $\PT$ symmetry is broken for any $\gamma$ because eigenvalues of  $H$ are complex for any  $\gamma>0$.
(We notice that the local $\PT$ symmetry can be fixed   if we add the diagonal matrix $\diag (1,-1,1,-1)$ to $H$.
In this case, both networks shown in fig.~\ref{fig08} have equal spectra~\cite{ZK}).

Operator $H$ is invertible unless $\mu =\pm\sqrt{1/2-\gamma^2}$ or $\mu = \gamma =0$.
Existence of analytical continuation of the one-quadrimer state from $\epsilon=0$
can be shown using the same ideas as the presented above. The only essential difference
is that now $\PT$-symmetric reduction is as follows:  at the central quadrimer,
we set  $w_0^{(1)} = -{\bar w_0^{(4)}}$ and  $w_0^{(2)} = {\bar w_0^{(3)}}$, while for $n\ne 0$,
we set $w_n^{(1)} = -{\bar w_{-n}^{(4)}}$ and  $w_n^{(2)} = { \bar w_{-n}^{(3)}}$.

Because local  $\PT$ symmetry is broken  for any $\gamma$,  the global  $\PT$ symmetry
of the infinite network is also broken  for small $\epsilon$.
Therefore, all the solitons bifurcating from the anticontinuum limit are unstable at least
for sufficiently small $\epsilon$. However, by increasing the coupling
parameter $\epsilon$, the global $\PT$ symmetry  is restored and the network in fig.~\ref{fig08}(b)
may possess stable solitons. An example of a stable discrete soliton for this network is shown
in  fig.~\ref{fig10}(b).

\section{Conclusion}
In this Letter we have shown that the idea of analytical continuation
from the anticontinuum limit  can be extended to the networks of   $\PT$-symmetric clusters,
offering analytical proof of the existence of localized discrete solitons.
Such solitons obey the $\PT$-symmetric shape and can be found stable.
As particular examples, we considered in details the chains of $\PT$-symmetric dimers
and the networks of $\PT$-symmetric quadrimers.

The considered systems allow
for further straightforward generalizations, say to chains of
clusters where there exist more than one link among the neighbor
ones, like the chain of dimers with pairwise coupling
considered in~\cite{Suchkov} or the chain of oligomers,
i.e. clusters with more than four sites.
Furthermore, the  approach of continuation from the anticontinuum limit
can be used for  developing of a  classification of intrinsic localized modes,
as well as analytical theory of the nonlinear stability of such modes.

\acknowledgments VVK and DAZ acknowledge support of the   FCT
(Portugal) grants: SFRH / BPD / 64835 / 2009, PTDC / FIS / 112624 / 2009, and
PEst-OE / FIS / UI0618 / 2011.

\end{document}